\documentclass[conference]{IEEEtran}
\IEEEoverridecommandlockouts

\usepackage{xcolor}
\definecolor{Gray}{gray}{0.9}
\usepackage[utf8]{inputenc} 
\usepackage{tabularx}
\usepackage{stfloats}

\newcommand{\review}[1]{{\color{black}{#1}}}

\usepackage{booktabs}
\usepackage[mode=text]{siunitx}[=2021-04-09]
\usepackage{multirow}
\usepackage{graphicx}
\usepackage[most]{tcolorbox}
\usepackage{arydshln}

\usepackage{textgreek}
\usepackage{makecell}
\usepackage{microtype}

\definecolor{vlightgray}{gray}{0.9}

\newcolumntype{L}[1]{>{\raggedright\let\newline\\\arraybackslash\hspace{0pt}}m{#1}}
\newcolumntype{C}[1]{>{\centering\let\newline\\\arraybackslash\hspace{0pt}}m{#1}}
\newcolumntype{R}[1]{>{\raggedleft\let\newline\\\arraybackslash\hspace{0pt}}m{#1}}

\newcommand\hypothesis[2]{\vspace{0.5em} \noindent \hangindent=1em \textbf{Hypothesis #1 (H#1).} {#2}\vspace{0.5em}}

\definecolor{mypurple}{rgb}{139,0,139}
\definecolor{mygreen}{rgb}{0,140,0}
\definecolor{orange}{RGB}{255,127,0}
\ifCLASSINFOpdf
\else
\fi

\usepackage{hyperref}

\hypersetup{
 colorlinks=true,  
 citecolor=blue,
 urlcolor=blue,
 linkcolor=blue,
 bookmarksnumbered=true,   
 bookmarksopen=true,     
 bookmarksopenlevel=1,     
 pdfstartview=Fit,      
 pdfpagemode=UseOutlines,  
 pdfpagelayout=TwoPageRight 
}

\begin{document}



\title{A Model for Understanding and Reducing Developer Burnout}



\author{

\IEEEauthorblockN{Bianca Trinkenreich\IEEEauthorrefmark{1}, Klaas-Jan Stol\IEEEauthorrefmark{2}, Igor Steinmacher\IEEEauthorrefmark{3}, Marco A. Gerosa\IEEEauthorrefmark{3},\\ Anita Sarma\IEEEauthorrefmark{4}, Marcelo Lara\IEEEauthorrefmark{1}, Michael Feathers\IEEEauthorrefmark{1}, Nicholas Ross\IEEEauthorrefmark{1}, Kevin Bishop\IEEEauthorrefmark{1}\\}
\\
\IEEEauthorblockA{\IEEEauthorrefmark{1}Globant, United States, \{first.lastname\}@globant.com}
\IEEEauthorblockA{\IEEEauthorrefmark{2}Lero and University College Cork, Ireland, k.stol@ucc.ie}
\IEEEauthorblockA{\IEEEauthorrefmark{3}Northern Arizona Uniersity, United States, \{igor.steinmacher, marco.gerosa\}@nau.edu}
\IEEEauthorblockA{\IEEEauthorrefmark{4}Oregon State University, United States, \{anita.sarma@oregonstate.edu}
}

\maketitle

\thispagestyle{plain}
\pagestyle{plain}
\begin{abstract}
Job burnout is a type of work-related stress associated with a state of physical or emotional exhaustion that also involves a sense of reduced accomplishment and loss of personal identity. Burnt out can affect one's physical and mental health and has become a leading industry concern and can result in high workforce turnover.
Through an empirical study at Globant, a large multi-national company, we created a theoretical model to evaluate the complex interplay among organizational culture, work satisfaction, and team climate, and how they impact developer burnout. 
We conducted a survey of developers in software delivery teams (n=3,281) to test our model and analyzed the data using structural equation modeling, moderation, and multi-group analysis. Our results show that Organizational Culture, Climate for Learning, Sense of Belonging, and Inclusiveness are positively associated with Work Satisfaction, which in turn is associated with Reduced Burnout. 
Our model generated through a large-scale survey can guide organizations in how to reduce workforce burnout by creating a climate for learning, inclusiveness in teams, and a generative organizational culture where new ideas are welcome, information is actively sought and bad news can be shared without fear.

\end{abstract}


\section{Introduction}
\label{intro}
Developers' well-being and work satisfaction have a strong influence on workforce retention~\cite{madigan2021towards,mullen2018job}. 
When organizations invest in the health and safety of its workforce, it is linked to organizational commitment among employees \cite{mearns2010investment} and results in returns that is 2x the amount invested \cite{phillips2014measuring,unsal2021return}.
On the other hand, employee attrition has significant costs, including disruption of ongoing working in a team as well as costs involved in recruiting and training a new team member.
This is particularly important for the software industry where  `job-hopping' is quite normal, with many developers changing jobs every few years \cite{gartner2021}. Developer retention is, therefore, a key to the long-term success of software organizations. 

Prior research in other fields suggests that burnout is an important factor in employees' intention to leave their job \cite{weisberg1994measuring}. Burnout refers to an individual's experiences of exhaustion on physical, emotional, and cognitive levels \cite{pines1981burnout}. Freudenberger was among the first to explore this concept, invoking a dictionary definition as \textit{``to fail, wear out, or become exhausted by making excessive demands on energy, strength, or resources''} \cite{Freudenberger1974}. While there has been considerable attention in the software engineering literature for themes such as job satisfaction \cite{francca2018motivation,sharma2020exploring,stol2022gamification}, there is a surprising paucity of research on burnout. Job burnout has become increasingly relevant in today's discussions on retaining talent. The COVID-19 pandemic caused a major shift in working patterns for knowledge workers starting in March 2020. Many developers felt overwhelmed working from home while also needing to take care of family and children. Others missed human contact with colleagues and support structures available in the office. As the pandemic has started to wind down (at the time of writing), scholars have coined the term ``Great Resignation'' to refer to initial observations that many workers across a variety of domains are voluntarily resigning from their job; one explanation is that the pandemic has triggered people to rethink their goals and ambitions in life \cite{serenko2022great}. As in other fields \cite{Sheather2021}, burnout is also likely playing a role in IT staff's decisions to leave an organization.

It is important, therefore, to understand what \textit{causes} burnout and factors that can mitigate it. 
\review{Following prior studies \cite{forsgren2016role,dora2019effect}, we look at how organizational culture relates to burnout. In particular,} we unpack this relationship by investigating a number of salient themes that have attracted interest in recent years, including employees' sense of belonging and work satisfaction.

Our goal is to identify the organizational and cultural antecedents that can reduce burnout. To achieve our goal, we defined the following research questions:\\[.1in]
\noindent \textit{\textbf{RQ1.} How are organizational culture and burnout related?}

\vspace*{2mm}

\noindent \textit{\textbf{RQ2.} Does the relationship between organizational culture and burnout vary by gender and leadership position?}\\

We answer these questions within the context of software delivery teams in Globant, a large company employing 25,000 people, and a global presence in 36 cities in 17 countries across five continents, which provides services in digital transformation and assisting IT organizations in automation. Globant invests in continuous training of its talent pool on technical and social skills and has several initiatives in place to retain talent and avoid attrition. Globant places the well-being of its employees at the forefront, investing in research to identify and proactively implement strategies to reduce employee burnout and attrition.

\begin{figure*}[!hb]
\centering
\includegraphics[width=0.7\textwidth]{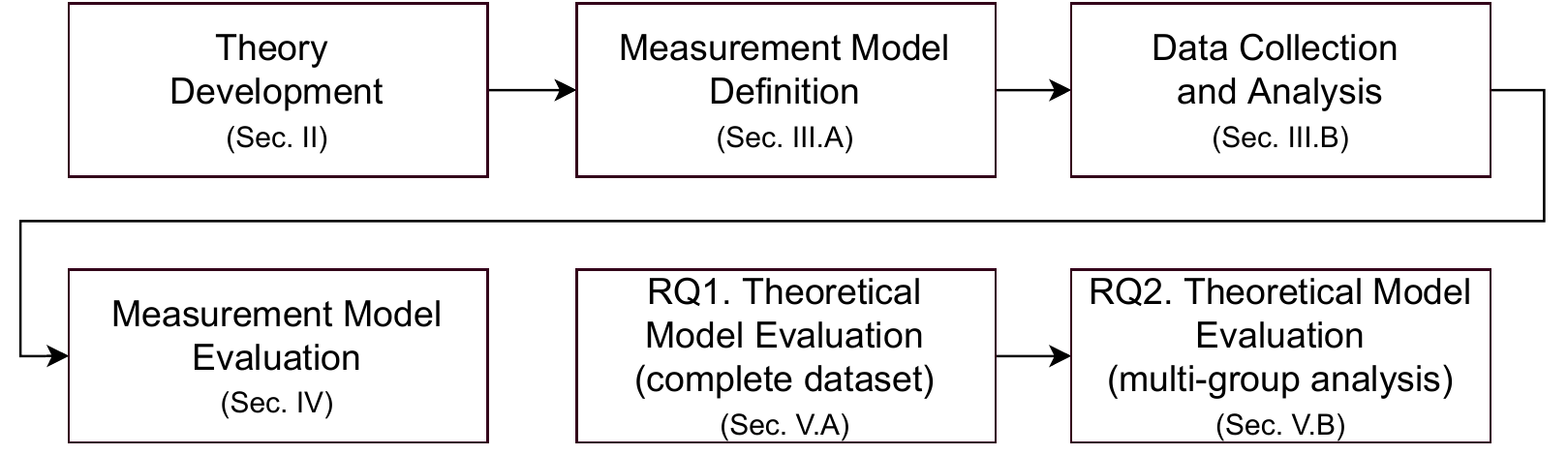}
\caption{Methodological approach}
\label{fig:research_design}
\end{figure*}

To answer our research questions, we developed a theoretical model of factors associated with burnout grounded in prior literature and tested it using structural equation modeling, moderation, and multi-group analysis. We tested the model with data collected via an online questionnaire (n=3,281) for current members of software delivery teams who work on different projects at Globant.
Fig.~\ref{fig:research_design} summarizes the study design.

Our results show that Organizational Culture, Climate for Learning, Sense of Belonging, and Inclusiveness are positively associated with Work Satisfaction, which in turn is associated with Reduced Burnout. 
A Climate for Learning improves Work Satisfaction for employees who do not hold leadership positions as compared to leadership. 
Team inclusiveness is positively associated with work satisfaction and has a bigger impact on women. Women are 2x more satisfied and less burnt out when their team is inclusive. 
National culture also plays a role between work satisfaction and burnout. Living in a masculine (and more competitive) culture further helps reduce burnout when men have work satisfaction; national culture does not play a role for women. An understanding of how these factors interplay can help organizations create a welcoming environment that improves developers' well-being and reduces workforce attrition. 

\section{Background and Hypothesis Development}
\label{sec:theory_development}

We review prior work related to this study and develop a theoretical model that reflects the interests of Globant managers.

\subsection{The Role of Organizational Culture in Sense of Belonging, Climate for Learning, and Inclusiveness}

An organization's culture affects people's daily work activities.
Organizational culture has been shown to influence software delivery performance \cite{dora,forsgren2016role}, staff well-being, and retention \cite{dora2019effect}, while also enticing software developers to support the company's business \cite{moreira2022organizational}. 
Westrum developed a typology of organizational cultures based on human factors in system safety, particularly in the context of accidents in technological domains, such as aviation and healthcare \cite{westrum2004typology}. The typology defines three types of organizations \review{in terms of information flow and psychological safety}. Pathological organizations exhibit low levels of cooperation across groups and a culture of blame. Bureaucratic cultures emphasize rules and positions and compartmentalize responsibilities by departments. Generative organizations are performance-oriented, with good information flow, high levels of cooperation and trust, and bridging between teams. The generative level can be achieved by creating cross-functional teams to improve cooperation, holding blameless postmortems, sharing risks and responsibilities, breaking down silos, and encouraging bridging, experimentation, and novelty.
An organizational culture where members of the team cooperate with each other and share responsibilities \cite{westrum2004typology} creates feelings of membership or being part of a team \cite{jelphs2016working}. This organizational culture presents the organization as an extended family, leading employees to develop a strong sense of belonging to the organization \cite{janicijevic2018influence}. Hence, we posit:

\hypothesis{1}{A Generative Organizational Culture has a positive association with a Sense of Belonging.}

An organization that exhibits a climate for learning makes resources available for continued education and offers continuous encouragement to teams to learn by providing them space and time to acquire new knowledge and explore ideas \cite{dora}. Organizational culture fosters the process of learning \cite{freiling2010organizational}. When holding blameless retrospectives and having out-of-box thinking, a generative organizational culture \cite{Wolbling2012} creates a positive Climate for Learning \cite{lok2004effect} as instead of punishing, the team is trained to learn from failures. Thus, our second hypothesis is:

\hypothesis{2}{A Generative Organizational Culture exhibits a Climate for Learning.}

When welcoming new ideas, a generative culture brings a positive tone to a welcoming space and a spirit of friendliness that leads to feelings of inclusiveness among the members of a team \cite{jelphs2016working}. When engaging in organizational culture, team members perceive an inclusive climate that leads to increased work satisfaction \cite{brimhall2019inclusion}. This leads to our third hypothesis:

\hypothesis{3}{A Generative Organizational Culture exhibits Inclusiveness.}

\subsection{The Role of Sense of Belonging, Climate for Learning and Inclusiveness in Work Satisfaction}

The need to belong is a powerful, fundamental, and pervasive force that has multiple strong effects on emotional patterns and cognitive processes, across all cultures and different types of people \cite{baumeister2017need}. Maslow \cite{maslow1943theory} positioned `belonging' as a basic human need, and Hagerty et al. \cite{hagerty1995developing} posited that a Sense of Belonging represents a unique mental health concept. A sense of belonging is key to work satisfaction \cite{lim2008job}, and productivity \cite{baumeister2017need}, and can help to avoid attrition \cite{allen2019making}. References to the importance of a sense of belonging are found throughout the psychological, health care, and education literature. On the other hand, a lack of a sense of belonging is linked to a variety of ill effects on health, adjustment, and well-being \cite{baumeister2017need}. Hence, we propose our fourth hypothesis:

\hypothesis{4}{Sense of belonging has a positive association with Work Satisfaction.}

Prior research on software delivery teams has shown that learning is associated with Work Satisfaction for software delivery teams, as learning is a valuable investment into the project’s future and also into the employee's own career \cite{meyer2019today}. Moreover, satisfying an employee's need for growth requires that the employee is satisfied with the opportunities to learn and advance at work \cite{janicijevic2018influence}. Thus we propose:

\hypothesis{5}{Climate for Learning has a positive association with Work Satisfaction.}

When feeling included by the team, employees believe they are valued for their unique personal characteristics and recognized as important members of the organization \cite{brimhall2019inclusion}. A perception of being socially included improves an individual's well-being \cite{jansen2016colorblind} and enhances their self-esteem and work satisfaction \cite{jansen2014inclusion}.
So, our sixth hypothesis is:

\hypothesis{6}{Inclusiveness has a positive association with Work Satisfaction.}




\subsection{The Role of Work Satisfaction in Reducing Burnout}

A decline in work satisfaction could signal burnout \cite{forsgren2021space}. Indeed, previous research showed that Burnout has an inverse relationship with Work Satisfaction \cite{dolan1987relationship,brewer2002burnout}. Thus, we propose:

\hypothesis{7}{Work Satisfaction has a reverse association with Burnout.}

\subsection{The Moderating Role of National Cultural Values on the association between Work Satisfaction and Burnout}
\label{sub:culture}

An individual’s response to stress is embedded within cultural beliefs. Cultural values are being accredited with a prominent role in various work-related predictor-outcome relationships, such as satisfaction, burnout \cite{rattrie2020culture}, and turnover \cite{sturman2012effect}. Globant has geographically distributed teams and needs to mitigate the social-derived challenges that are inherent in cultural differences. There are various classifications attempting to quantify cultural values such as the work by Hofstede \cite{hofstede2001culture}, Schwartz \cite{schwartz1999theory}, and the GLOBE study \cite{chhokar2007culture}. In this study, we adopt Hofstede’s classification, which was previously used to analyze the culture of software engineers \cite{lambiase2022good} and to investigate burnout \cite{rattrie2020culture}. Hofstede \cite{hofstede2001culture} defined the Hofstede’s 6-D framework with the following six dimensions of culture per country that assume values from zero to one hundred \cite{hofstede2011dimensionalizing}: 

\textit{Power Distance} refers to authority and hierarchy and expresses the degree to which less powerful members of a society accept and expect that power is distributed unequally. High power distance means an acceptance of hierarchical order in which people have a determined place. Low power distance means a desire for an egalitarian distribution of power \cite{hofstede2011dimensionalizing,hofstede2005cultures}. In high power distance cultures, social hierarchy is established and executed clearly and without reason \cite{gokmen2021impact}. Hierarchy in an organization is seen as reflecting inherent inequalities, centralization is popular, subordinates expect to be told what to do, and the ideal boss is a benevolent autocrat \cite{hofstede2011dimensionalizing,hofstede2005cultures}. 
In Hofstede's classification \cite{hofstede2011dimensionalizing}, Mexico and India are examples of hierarchical societies with high Power Distance. 



\textit{Individualism} represents the degree to which people in a society are integrated into groups. High individualism indicates people who take care of only themselves and their immediate families and should not rely (too much) on authorities for support. In contrast, low individualism (collectivism) reflects a closer integration into cohesive in-groups in which people protect each other with unquestioning loyalty \cite{hofstede2011dimensionalizing,hofstede2005cultures}. Collectivists mostly pursue group goals and improve group level engagement \cite{ronen2009attachment}. 
In Hofstede's classification \cite{hofstede2011dimensionalizing}, the United States is an example of a society with high Individualism.

\textit{Masculinity} is defined as a preference for achievement, heroism, assertiveness, and material rewards for success. While high masculinity societies are materialist and competitive, low masculinity culture (femininity) is more cooperative, consensus-oriented, caring for the weak, and prevailing the life quality \cite{hofstede2011dimensionalizing,hofstede2005cultures}. Japan is an example of high degree of masculinity in Hofstede's classification \cite{hofstede2011dimensionalizing}.

\textit{Uncertainty Avoidance} expresses the degree to which people keep away from ambiguity. Cultures high in uncertainty avoidance tend to focus on rules, structured activities, employee security, and stability. Low levels of uncertainty avoidance have a more relaxed attitude in which practice is more important than rules \cite{hofstede2011dimensionalizing,hofstede2005cultures}. In Hofstede's classification \cite{hofstede2011dimensionalizing}, Uruguay is an example of a score in this dimension while China is the opposite.

\textit{Long Term Orientation} measures the degree a culture will keep some links with its own past while dealing with the challenges of the present and the future. A high degree in this index indicates more pragmatic people who have perseverance and patience to prepare for the future. On the contrary, a lower value on this index (indicating short-term orientation) indicates that people have a more narrow-minded focus and sensitivity to immediate outcomes of their actions, tending to value steadfastness, and considering a societal change with suspicion \cite{hofstede2011dimensionalizing,hofstede2005cultures}. People in such societies have a strong concern with establishing the absolute truth; they are normative in their thinking, exhibit great respect for traditions, a relatively small propensity to save for the future, and a focus on achieving quick results.
Argentina, for example, has a high score in this dimension in Hofstede's classification \cite{hofstede2011dimensionalizing}.

\textit{Indulgence} is related to the degree of freedom the societal norms give citizens to fulfill human desires. A high degree indicates a society that relatively allows free gratification of basic and natural human desires related to hedonism. Conversely, low levels of indulgence (restraint) indicate a society that controls gratification of needs and regulates it using strict social norms \cite{hofstede2011dimensionalizing,hofstede2005cultures}.

\begin{figure*}[!hb]
\centering
\includegraphics[width=0.7\textwidth]{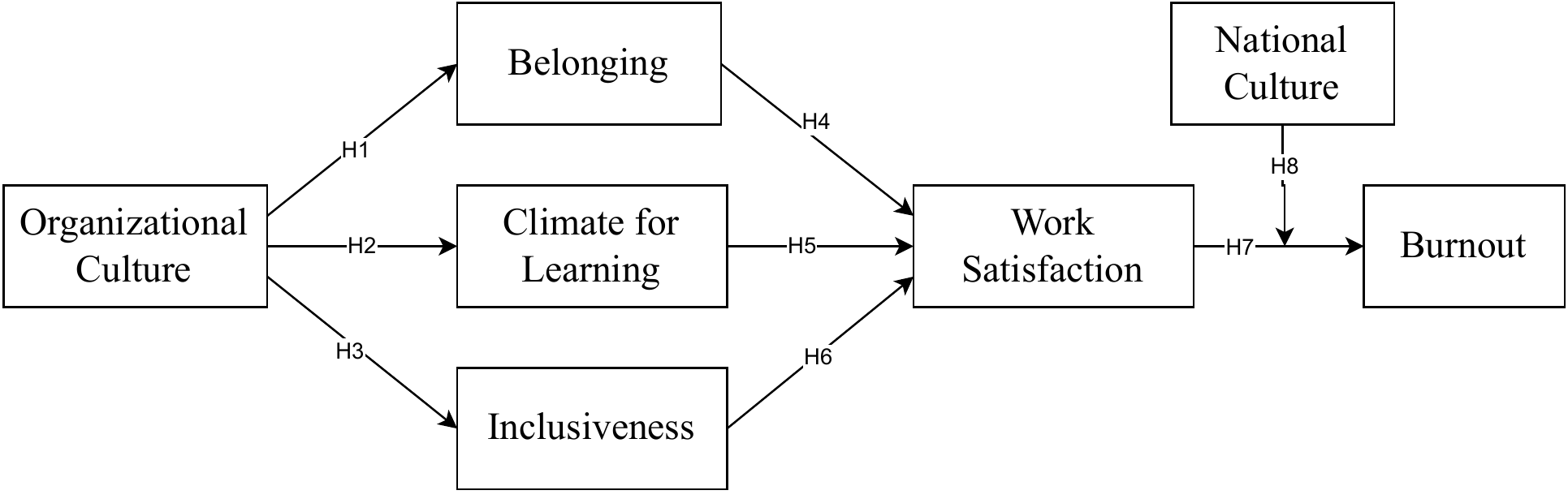}
\caption{Theoretical model}
\label{fig:theoretical_model}
\end{figure*}

Studies from two decades ago showed that Long Term Orientation \cite{jung1995bridging} and Power distance \cite{lin2013abusive} could help foster organizational well-being. 
\review{Subordinates who are surrounded by a  high power distance cultural value evaluate abusive supervision as irrelevant to their well-being \cite{lin2013abusive}.}
Individualism could reduce well-being \cite{triandis2000cultural}, as unpleasant life events are not met with sufficient social support. In workplace contexts, managers face increasingly complex and subtle differences among employees that reflect cultural influences from the country's culture. Thus, we argue that country culture moderates the link between Work Satisfaction and Burnout. Conceptually, we model this as a single moderator (see Fig.~\ref{fig:theoretical_model}), but we propose that each of Hofstede's six dimensions has such a moderating role. Thus, we propose:


\hypothesis{8}{(a) Power distance, (b) individualism, (c) masculinity, (d) uncertainty avoidance, (e) long-term orientation, and (f) indulgence moderate the effect of Work Satisfaction on Burnout.}

Fig.~\ref{fig:theoretical_model} presents the theoretical model.

\section{Research Design}
Management at Globant was keen to understand the relationships that we proposed in Fig.~\ref{fig:theoretical_model} and obtain an answer to RQ1, which develops an understanding of how organizational culture and burnout in software delivery teams are related. To evaluate the model, we conducted a survey among software delivery team members at Global and analyzed the data using Partial Least Squares (PLS) Structural Equation Modeling (SEM) \cite{ringle2015structural}. SEM facilitates the simultaneous analysis of relationships among constructs, each measured by one or more indicator variables. 
Then, to answer RQ2, we used a Multi-Group Analysis (MGA) to establish whether these relationships vary by gender and leadership position.

A recent survey of the use of PLS in software engineering (which also provides an introduction to PLS) revealed that PLS-SEM has been used to study a variety of phenomena in software engineering \cite{russo2021pls}. 
For example, it has previously been used to study job satisfaction and turnover intentions of software engineering teams \cite{sharma2020exploring} and the success factors of a large-scale Agile Software transformation process \cite{russo2021pls}.

In this section, we discuss the measurement model (how each theoretical construct was measured) and the data collection and analysis. 

\subsection{Measurement model}
\label{sec:measurement_model}

The theoretical model comprising the hypotheses is based on a number of theoretical concepts; some of the concepts cannot be directly observed (e.g., Climate for Learning, Organizational Culture, and Work Satisfaction)---these concepts are represented as \textit{latent} variables. A latent variable cannot be directly measured or observed, but instead is measured through a set of indicators or manifest variables. For the latent variables in this study, we adapted existing measurement instruments when possible.
We define the constructs below and list the complete questionnaire in the replication package \cite{replication_package}.


\textbf{Organization Culture} was measured as a latent construct represented by six five-point Likert questions. The questions were adapted from the Westrum Culture \cite{westrum2004typology}, which has previously been used as an instrument to measure organizational culture in software delivery teams \cite{forsgren2018accelerate,forsgren2016role}.

\textbf{Belonging} was measured using a five-point Likert question to assess aspects of membership, as being part of the team.

\textbf{Climate for Learning} was measured as a latent construct represented by two five-point Likert questions. The questions were inspired by DORA Research Program \cite{dora} and designed to evaluate if members of the team perceive that the team considers learning as an investment rather than a cost, and essential for continued progress.

\textbf{Inclusiveness} was measured through one question to assess if the team has a safe space for diversity in which everyone is welcomed and treated equally and fairly.

\textbf{Work Satisfaction} was measured as a latent construct composed of two Employee Net Promoter 10-scale Score (eNPS) questions \cite{reichheld2006ultimate,sedlak2020employee} and one five-point Likert question about enjoyment. The eNPS questions were towards the team and towards the company, which relies on asking the respondent the willingness to recommend the team and the company to friends and colleagues.

\textbf{Burnout} was measured as a latent construct composed of two five-point Likert items. 
While instruments exist to measure burnout, these are very long, which would result in an overly long survey instrument. We took a pragmatic approach and focused on two statements: (1) the extent to which a team has a manageable workload with sustainable levels of stress, and burnout is not perceived as a significant problem or risk; and (2) the extent to which tasks are assigned in a way that allows enough time to achieve commitments, and team members are able to focus on one process at a time. Both were measured as `reversed' items, i.e., a strong \textit{disagreement} indicated a higher level of burnout.


\textbf{National Culture:} Based on the respondents' country of residence, we used Hofstede’s classification of National Culture as moderators (Sec.~\ref{sub:culture}). This classification's six-dimensional approach to cultural variation includes power distance, individualism/collectivism, masculinity/femininity, uncertainty avoidance, long-term/short-term orientation, and indulgence \cite{hofstede2001culture}.

\subsection{Data Collection and Analysis}
\label{sec:data_collection_and_analysis}

We administered an online questionnaire using Globant Glow,\footnote{\url{https://os.starmeup.com/en.html}} which was answered by members of software delivery teams at Globant.
%

%
The survey was sent to respondents by email using a corporate address. The leader of each team encouraged team members to fill the questionnaire out during regular meetings.
We received 10,566 responses; however, our analysis techniques require complete responses, and we removed 7,285 responses that contained blanks. Our final sample size has 3,281 responses. Table~\ref{tab:demographics} presents a summary of the respondents' characteristics.

\begin{table}[!t]
\centering
\caption{Demographics of respondents (n=3,281)}
\label{tab:demographics}
\sisetup{
    locale = US,
    mode=text,
    group-digits = integer,
    group-minimum-digits=4,
    group-separator={,},
    input-symbols = ( ) [ ] - + *,
    detect-weight=true, 
    detect-family=true,
    table-format=0.2,
    add-decimal-zero=false, 
    add-integer-zero=false,
    round-mode=places, 
    round-precision=1, 
    parse-numbers = true
}
\begin{tabular}{p{5cm}
                S[table-format=4.0]
                S[table-format=2.1]}
\toprule
Attribute & {N} & {Percentage}\\
\midrule
\multicolumn{3}{c}{Gender}\\
\midrule
Men & 2487 & 74.8\%\\
Women & 794 & 25.2\%\\

\midrule
\multicolumn{3}{c}{Country of Residence}\\
\midrule
Colombia & 789 & 24.0\% \\
Argentina & 721 & 22.0\% \\
India & 581 & 17.7\% \\
Mexico & 515 & 15.7\% \\
Uruguay & 211 & 6.4\% \\
Chile & 197 & 6.0\% \\
Peru & 128 & 3.9\% \\
USA & 55 & 1.7\% \\
Brazil & 53 & 1.6\% \\
Spain & 15 & 0.5\% \\
Belarus & 9 & 0.3\% \\
Others & 7 & 0.3\% \\

\midrule
\multicolumn{3}{c}{Roles}\\
\midrule
\multicolumn{3}{l}{Leadership Positions}\\
\midrule

Project Manager & 248 & 7.6\%\\
Tech Manager & 34 & 1.0\% \\
Product Manager & 12 & 0.4\% \\
Other leadership roles & 5 & 0.2\% \\

\midrule
\multicolumn{3}{l}{Non-Leadership Positions}\\
\midrule
Developers & 1723 & 52.5\%\\
Test/Quality Assurance & 656 & 20.0\%\\
Business Analyst/Intelligence & 158 & 4.8\%\\
ERP Tech/Functional & 98 & 3.0\%\\
DevOps Engineer & 71 & 2.2\%\\
Designer/Artist & 65 & 2.0\%\\
Data Architect/Scientist & 55 & 1.7\%\\
SysAdmin/Cloud Engineer & 45 & 1.3\%\\
Other non-leader roles & 111 & 3.3\% \\

\midrule
\multicolumn{3}{c}{Starting year at the company}\\
\midrule
Between 2021 and 2022 & 1460 & 44.5\%\\
Between 2019 and 2020 & 1081 & 32.9\%\\
Between 2018 and 2017 & 422 & 12.9\%\\
Between 2016 and 2015 & 181 & 5.5\%\\
Between 2014 and 2015 & 137 & 4.2\%\\

\bottomrule

\end{tabular}

\end{table}


We used SmartPLS version 4 for the analyses; SmartPLS is a proprietary package for analyzing PLS models. The analysis comprised three steps, with tests and procedures in each step. The first step was to evaluate the measurement model (Sec.~\ref{sec:results_measurement_model}), which empirically assesses the relationships between the constructs and indicators. The second step was to evaluate the theoretical model that represents the set of hypotheses (Sec.~\ref{sec:results_structural_model}). The third step was to evaluate observed heterogeneity by multi-group analysis of gender and leadership position (Sec.~\ref{sec:results_mga}). 


PLS does not make assumptions about the distribution (such as a Normal distribution) of the data; without knowing the distribution of the data, parametric tests (which are based on a distribution with certain parameters) cannot be used to establish the standard error and thus significance. Instead, PLS packages employ a `bootstrapping' procedure: it draws a large number (e.g. 5,000) of random `subsamples' of the same size as the original sample (using replacement). The model is estimated for each subsample, generating a sampling distribution, which is used to determine a standard error \cite{hair2019use}, which can subsequently be used to make statistical inferences. 
The mean path coefficient determined by bootstrapping can differ slightly from the path coefficient calculated directly from the sample. 

The online appendix provides results of a variety of additional analyses and validity checks \cite{replication_package}.

\section{Measurement Validity and Model Fit}
\label{sec:results_measurement_model}

\subsection{Measurement Validity}
As a first step, we conducted two recommended tests to ensure that a dataset is suitable for factor analysis, i.e., that the variables in a dataset can be reduced to a smaller number of factors \cite{bartlett1950tests,hair1995multivariate}. The first test is Bartlett’s test of sphericity \cite{bartlett1950tests} on all constructs. We found a p-value \textless{} .01 (p-values less than .05 indicate that factor analysis may be suitable). Second, we calculated the Kaiser-Meyer-Olkin (KMO) measure of sampling adequacy. Our result (.92) is well above the recommended threshold of .60 \cite{hair1995multivariate}.

Afterward, we conducted several tests to validate the measurement of our theoretical concepts, including convergent validity, internal consistency reliability, discriminant validity, and collinearity, as discussed next.

\subsubsection{Convergent Validity}

First, we assess the convergent validity of the measurement instrument, i.e., we assess whether the questions (indicators) that represent each latent variable are understood by the respondents in the same way as they were intended by the designers of the questions \cite{kock2014advanced}. This assessment relates to the degree to which a measure correlates positively with alternative measures of the same construct. 
Our model contains four latent variables (Climate for Learning, Organizational Culture, Work Satisfaction, and Burnout). Changes in the theoretical, latent construct should be `reflected' in changes in the indicator variables \cite{hair2019use}; for example, if 
Work Satisfaction increases, a concept we cannot measure or observe \textit{directly}, we expect to see this change reflected in the values of its indicators that we \textit{can} measure or observe directly.

We used two metrics to assess convergent validity: the Average Variance Extracted (AVE) and the loading of an indicator onto its construct (the outer loading). The AVE is the proportion of variance that is shared across indicators. 
The AVE should be at least 50\%, indicating that it explains most part the variation in its indicators \cite{hair2019use}. 
All AVE values for the three latent constructs in our model are above this threshold of 50\% (see appendix).

\begin{figure*}[!b]
\centering
\includegraphics[width=0.7\textwidth]{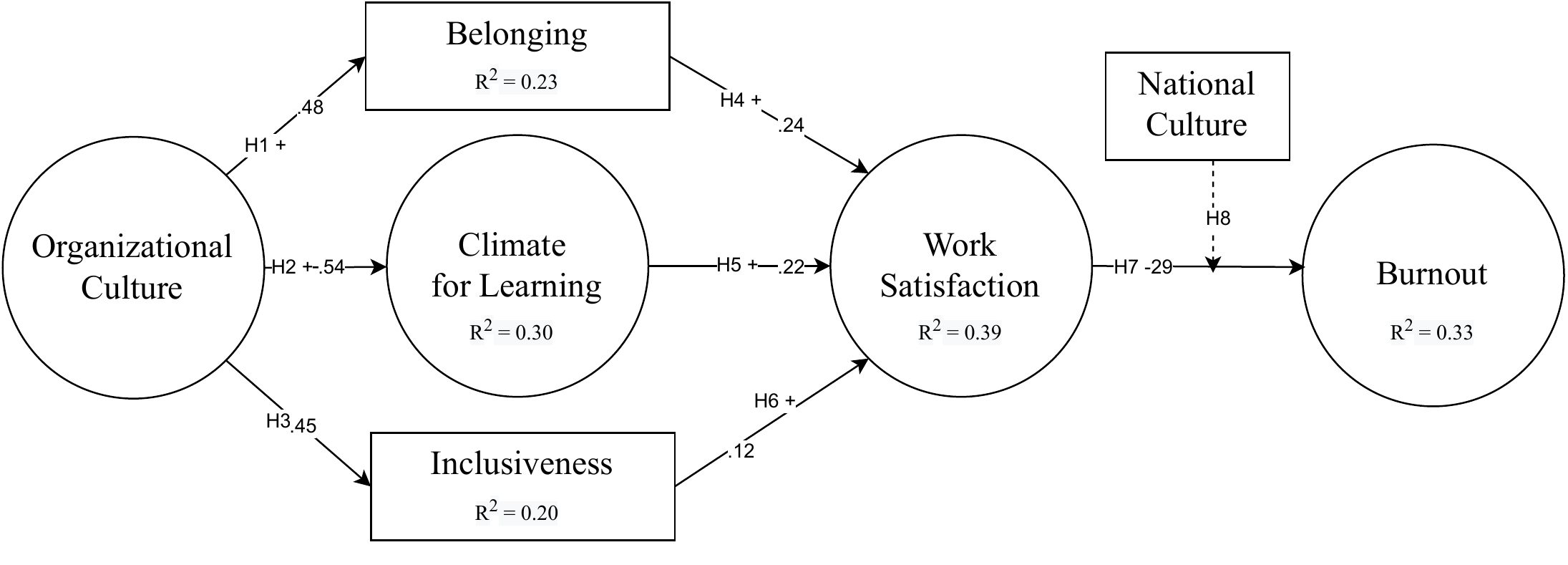}
\caption{Path coefficients (p $<$ 0.05 indicated by a full line). Latent constructs are represented as circles.}
\label{fig:evaluating_structutal_model}
\end{figure*}

A latent variable is measured by two or more indicators; 
each indicator is expected to have a loading of at least 50\%, because the square of the loading indicates the variance in the indicator that is explained, which should be at least 50\% (and 70\%\textsuperscript{2} $\approx$ 50\%) \cite{hair2019use}. All loadings of the indicators of all four latent constructs exceeded this as shown in Figure~\ref{fig:evaluating_structutal_model}, which we considered acceptable.

\subsubsection{Internal Consistency Reliability}

Second, we verified how well the different indicators are consistent with one another and able to reliably and consistently measure the constructs. 
A high degree of consistency means that indicators refer to the same construct. There are several tests to measure internal consistency reliability. We performed both the Cronbach's \textalpha{} and Composite Reliability tests; Cronbach's \textalpha{} frequently shows lower values, whereas the Composite Reliability (CR) is a more liberal test, which sometimes overestimates the values \cite{hair2019use}.
A desirable range of values for both Cronbach’s \textalpha{} and CR is between .7 and .9 \cite{hair2019use}. Values below .6 suggest a lack of internal consistency reliability, whereas values over .95 suggest that indicators are too similar and thus are not desirable. All Cronbach \textalpha{} and CR values fell between .7 and .9 (see appendix).



\subsubsection{Discriminant Validity}

Third, we verified whether each construct represents characteristics that are not measured by other constructs, i.e., we assessed the discriminant validity of the constructs. 
A primary means to assess discriminant validity is to investigate the Heterotrait-monotrait (HTMT) ratio of correlations \cite{henseler2015new}. The discriminant validity could be considered problematic if the HTMT ratio exceeds .9 \cite{henseler2015new}; some scholars recommend a more conservative cut-off of .85 \cite{hair2019use}. The HTMT ratio between the four latent constructs ranged between .65 and .71 (see appendix). 


\subsubsection{Assessing Collinearity}

To ensure that the variables are independent, we calculate their collinearity by means of the Variance Inflation Factor (VIF). In our model all VIF values are below 1.7, well below the cut-off of 5 \cite{hair2019use} (see appendix).

\subsection{Model Fit}
The overall model was measured through a standardized root mean square residual (SRMR) composite factor model, which should be lower than .08 \cite{hair2017advanced}. Thus, the values obtained for the complete model (.060), the men's model (.061), the women's model (.061), the leaders' model (.070), and the non-leaders model (.60) have a good fit.

\section{Results}
\label{sec:results}

To answer RQ1, we evaluated the hypotheses in the structural model (Sec.~\ref{sec:results_structural_model}) and, to answer RQ2, we performed a Multi-Group Analysis (Sec.~\ref{sec:results_mga}). 

\subsection{RQ1. How are organizational culture and burnout in software delivery teams related?}
\label{sec:results_structural_model}

Table~\ref{tab:new_path_analysis} shows the results for our hypotheses, including the mean of the bootstrap distribution (\textit{B}), the standard deviation (\textit{SD}), the 95\% confidence interval, and the p-values.

\review{Path coefficients (\textit{B}) are interpreted as follows in this example for H1: having \textit{B}=.48 means that a unit-change of Organizational Culture’s standard deviation triggers a direct change in Belonging of .48 $\times$ Belonging’s standard deviation.}

\begin{table}[!t]
\centering
\caption{Standarized path coefficients, standard deviations and confidence intervals. coefficients with * are \review{lower than .05}}
\label{tab:new_path_analysis}
\robustify{\bfseries}
\sisetup{
    mode=text,
    group-digits = false ,
    input-signs ={-},
    input-symbols = ( ) [ ] - + *,
    detect-weight=true, 
    detect-family=true,
    table-format=0.2,
    add-decimal-zero=false, 
    add-integer-zero=false,
    round-mode=places, 
    round-precision=2, 
    parse-numbers = true
}
\begin{tabular}{L{4.8cm}
                S
                S
                C{1.45cm}}
\toprule
& {\textit{B}} & {SD} & {95\% CI} \\
\midrule
\hangindent1em H1 Organizational Culture$\rightarrow$Belonging & .48* & .02 & (.44, .51)\\
\hangindent1emH2 Organizational Culture$\rightarrow$Climate for Learning & .54* & .02 & (.51, .57)\\
\hangindent1emH3 Organizational Culture$\rightarrow$Inclusiveness & .45* & .02 & (.41, .49)\\
\hangindent1emH4 Belonging$\rightarrow$Work Satisfaction & .24* & .02 & (.20, .27)\\
\hangindent1emH5 Climate for Learning$\rightarrow$Work Satisfaction & .22* & .02 & (.18, .25)\\
\hangindent1emH6 Inclusiveness$\rightarrow$Work Satisfaction & .12* & .02 & (.08, .16)\\
\hangindent1emH7 Work Satisfaction$\rightarrow$Burnout & -.29* & .02 & (\num{-.25}, \num{-.33})\\
\midrule
Moderators & & &\\
\midrule
\hangindent1emH8.a Power Distance \texttimes{} Work Satisfaction$\rightarrow$Burnout & .08 & .05 & (.19, .00)\\
\hangindent1emH8.b Individualism \texttimes{} Work Satisfaction$\rightarrow$Burnout & .05 & .04 & (.13, .03)\\
\hangindent1emH8.c Masculinity \texttimes{} Work Satisfaction$\rightarrow$Burnout & -.02 & .05 & (.07, \num{-.12})\\
\hangindent1emH8.d Uncertainty Avoidance \texttimes{} Work Satisfaction$\rightarrow$Burnout & -.01 & .05 & (.08, \num{-.01})\\
\hangindent1emH8.e Long Term Orientation \texttimes{} Work Satisfaction$\rightarrow$Burnout & -.09 & .08 & (.05, \num{-.27})\\
\hangindent1emH8.f Indulgence \texttimes{} Work Satisfaction$\rightarrow$Burnout & -.02 & .05 & (.08, \num{-.13})\\

\bottomrule

\end{tabular}
\end{table}


Results revealed that a Generative Organizational Culture has a positive significant association with Sense of Belonging (H1, \textit{B}$=$.48), Climate for Learning (H2, \textit{B}$=$.54), and Inclusiveness (H3, \textit{B}$=$.45). We also found that a Sense of Belonging to the team (H4, \textit{B}$=$.24), Climate for Learning (H5, \textit{B}$=$.22), and Inclusiveness (H6, \textit{B}$=$.12) have a positive and significant association with Work Satisfaction, which includes feelings of joy and enthusiasm to recommend the team and company as a place to work to friends former colleagues. Finally, Work Satisfaction has a reverse (negative) and significant association with Burnout (H7, \textit{B}=\num{-.29}). Hence, hypotheses H1 to H7 were supported with p-values $<$ 0.001.

We also investigated whether the association between Work Satisfaction and Burnout would change when considering  respondents' national culture, as defined by Hofstede's six dimensions (see Sec.~\ref{sub:culture}) (H8). This association is not significantly affected (neither increased nor reduced) by any of these six cultural dimensions. Power Distance (H8.a) and Individualism (H8.b) have a positive, but insignificant moderation effect in the association between Work Satisfaction and Burnout. Masculinity (H8.c), Uncertainty Avoidance (H8.d), Long Term Orientation (H8.e), and Indulgence (H8.f) have a negative, but insignificant moderation effect on the association between Work Satisfaction and Burnout. Hence, H8.(a-f) are not supported for the complete dataset.


We assessed the relationship between constructs and the predictive capabilities of the model. The \textit{R}\textsuperscript{2} ranges from 0 to 1, with higher values indicating a greater explanatory power. \textit{R}\textsuperscript{2} values of .75, .50, and .25 are considered substantial, moderate, and weak, respectively \cite{hair2019use}. However, such thresholds are rather arbitrary and generic, and do not consider the specific context or research area; in some fields, an \textit{R}\textsuperscript{2} value as low as .10 is considered satisfactory \cite{raithel2012value}.
The \textit{R}\textsuperscript{2} values of the five endogenous variables in our model (Belonging, Climate for Learning, Inclusiveness, Work Satisfaction, and Burnout) are shown in Table~\ref{tab:mga_gender} are acceptable ranging between .20 and .39 (column `All'; further variation in range emerged after the multi-group analysis, discussed later). 

We also assessed the predictive relevance of the model, using the Stone-Geisser \textit{Q}\textsuperscript{2} measure. For this, we used the PLSPredict algorithm that is available in the SmartPLS v. 4 package \cite{shmueli2016elephant,shmueli2019predictive}. Values larger than 0 indicate the construct has predictive relevance, while negative values (smaller than zero) indicate that the model does not perform better than the simple average of the endogenous variable would do. The values were all positive, indicating the construct has predictive relevance \cite{hair2019use}.

\subsection{RQ2: Does the relationship between organizational culture and burnout vary by gender and leadership position?}
\label{sec:results_mga}

RQ2 seeks to establish whether the theorized relationship between organizational culture and burnout (as investigated for RQ1), varies when we consider gender and leadership position. Are some of the hypothesized links stronger for men than for women, or vice versa? Or are these associations different for people in leadership positions vs. people not in leadership positions? To answer this, we used multi-group analyses splitting by gender and leadership position and exploring differences that can be traced back to observable characteristics and may not be evident when examined as a whole. The multi-group analysis involves running the PLS path model multiple times for different groups, once for each group; groups are captured through categorical variables (in this case, binary variables). 
Hair et al. \cite{hair2017advanced} proposed three steps to conduct such an analysis: (1) group creation; (2) invariance test; and (3) result analysis.

\subsubsection{Step 1. Groups Creation}
We grouped our participants to observe heterogeneity according to two variables: gender (male = 0 and female = 1) and leadership (leadership role = 1 and non-leadership role = 0). We used pre-existing demographic data the company maintains for its reporting requirements under government laws to split the participants into different groups. 



\subsubsection{Step 2. Evaluation of measurement invariance of composite models (MICOM)}
Measurement invariance is a mechanism to assess whether or not the loadings of the items that represent the latent variables differ significantly across different groups.
In other words, we want to assess whether the differences can be attributed to the theoretical constructs and not to how we measured those constructs \cite{hair2017advanced}. Comparing group-specific model relationships for significant differences using a multi-group analysis requires establishing configural and compositional invariance \cite{henseler2016testing,hair2017advanced}. Configural invariance does not include a test and is a qualitative assessment of making sure that all of the composites are equally defined for all of the groups such as equivalent indicators per measurement model, equivalent treatment of the data, and equivalent algorithm settings or optimization criteria. The configural invariance is established in our model. Following that, compositional invariance exists when the composite scores are the same across both groups, and is statistically tested to assess whether the composite scores differ significantly across the groups. For this purpose, the MICOM procedure examines the correlation between the composite scores of both groups and requires that the correlation equals 1. We ran the permutation test in SmartPLS and verified that compositional invariance is established for all latent variables in the PLS path model. We established partial measurement invariance and thus multi-group analysis is suitable \cite{ringle2016gain}.

\subsubsection{Step 3. Groups Comparison and Analysis}
Path coefficients generated from different samples are usually numerically different, but the question is whether the differences are statistically significant. We analyzed the differences between the coefficients' paths for the groups. If they are significant, they can be interpreted as having moderating effects. 


\textbf{Gender:} As Table~\ref{tab:mga_gender} shows, Generative Organizational Culture has a strong and significant relationship with Sense of Belonging and Climate for Learning for both men and women. However, although Organizational Culture is also associated with Inclusiveness for both men and women, the association is stronger for women (\textbeta{} = .53) than for men (\textbeta{} = .41). Sense of Belonging and Climate for Learning have a significant and similar relationship with Work Satisfaction for both genders. Although both genders are satisfied by Inclusiveness, women (\textbeta{} = .20) are two times more satisfied when the team is Inclusive compared to men (\textbeta{} = .10). Lastly, the link between Work Satisfaction and Burnout is the same for men and women.
However, men (but not women) who live in a competitive national culture, where people want to be the best (i.e., high degree of Masculinity), have even less burnout.

\begin{table*}[!ht]
\centering
\caption{Multi-Group Analysis: coefficients marked with $*$ are statistically significant; coefficients set in boldface indicate that the difference between groups (i.e. male vs. female, and non-leadership vs. leadership roles) is statistically significant}
\label{tab:mga_gender}
\robustify{\bfseries}
\sisetup{
    mode=text,
    input-signs ={-},
    group-minimum-digits=4,
    group-digits=true,
    group-separator = {,},
    input-symbols = ( ) [ ] - + *,
    detect-weight=true, 
    detect-family=true,
    add-decimal-zero=false, 
    add-integer-zero=false,
    round-mode=places, 
    round-precision=2, 
    parse-numbers = true
}
\begin{tabular}{L{8.2cm}
                S[table-format=4.2]
                S[table-format=2.2]
                p{5mm}
                S[table-format=4.2]
                S[table-format=2.2]
                p{5mm}
                S[table-format=4.2]}

\toprule

& \multicolumn{5}{c}{Multi-Group Analysis}\\
\cmidrule{2-6}
& \multicolumn{2}{c}{Gender} && \multicolumn{2}{c}{Leadership} & \\
\cmidrule{2-3}\cmidrule{5-6}
& {Male} & {Female} && {Non-leaders} & {Leaders} && {All}\\

\midrule
Sample size (N) & 2487 & 79 && 2982 & 299 && 3281\\
\midrule 

Belonging (\textit{R}\textsuperscript{2}) & .23 & .24 && .22 & .25 &&  .23\\
Climate for Learning (\textit{R}\textsuperscript{2}) & .28 & .32 && .29  & .31  && .29\\

Inclusiveness (\textit{R}\textsuperscript{2}) & .17 & .28 && .20 & .25 && .20\\

Work Satisfaction (\textit{R}\textsuperscript{2}) & .38 & .41 && .39 & .35 && .39\\

Burnout (\textit{R}\textsuperscript{2}) & .31 & .37 && .34 & .36 && .33\\

\midrule 
\hangindent1em H1 Organizational Culture $\rightarrow$ Belonging & .48* & .49*  && .47* &  .50* && .48*\\ 

\hangindent1em H2 Organizational Culture $\rightarrow$ Climate for Learning & .53*  & .57* && .54* & .56* && .54*\\ 

\hangindent1em H3 Organizational Culture $\rightarrow$ Inclusiveness & \bfseries .41* & \bfseries .53* && .44* & .46* && .45*\\

\hangindent1em H4 Belonging $\rightarrow$ Work Satisfaction & .25* & .21* && .24* & .20*  && .24* \\ 

\hangindent1em H5 Climate for Learning $\rightarrow$ Work Satisfaction  & .22* & .21* && \bfseries .22* & \bfseries.08 && .22*\\

\hangindent1em H6 Inclusiveness$\rightarrow$Work Satisfaction & \bfseries .10* & \bfseries .20* && .13* & .01* && .12*\\ 

\hangindent1em H7 Work Satisfaction$\rightarrow$Burnout & -.29* & -.29* && -.29* & -.20* && -.29*\\
\midrule
Moderators\\

\midrule

\hangindent1em H8.a Power Distance \texttimes{} Work Satisfaction$\rightarrow$ Burnout & .01 & .01 && .11 & .07 && .08\\

\hangindent1em H8.b Individualism \texttimes{} Work Satisfaction$\rightarrow$ Burnout & .08 & -.01 && .16 & .03 && .05\\

\hangindent1em H8.c Masculinity \texttimes{} Work Satisfaction $\rightarrow$ Burnout & \bfseries -.10* & \bfseries .13 && -.17 & .01 && -.02\\

\hangindent1em H8.d Uncertainty Avoidance \texttimes{} Work Satisfaction $\rightarrow$ Burnout & -.07 & .18 && -.01 & .01 && -.01\\

\hangindent1em H8.e Long Term Orientation\texttimes{} Work Satisfaction$\rightarrow$  Burnout & -.13 & -.04 && -.10 & -.06 &&  -.09\\

\hangindent1em H8.f Indulgence \texttimes{} Work Satisfaction $\rightarrow$ Burnout & .03 & -.18 && .15 & -.04 && -.02\\
\bottomrule

\end{tabular}

\end{table*}

\textbf{Leadership Position:} As Table~\ref{tab:mga_gender} shows, Climate for Learning has a strong and significant relationship with Work Satisfaction for those who are not in leadership positions. However, Climate for Learning is not associated with Work Satisfaction for leaders. Lastly, leaders (\textbeta{} = .41) are close to two times more satisfied by a Generative Organizational Culture when compared to those who are not in leadership positions (\textbeta{} = .24).

\section{Discussion}

Human factors are receiving increasingly more attention in software engineering research and industry. Themes such as work satisfaction have been studied extensively and have been linked to employees' intention to stay with (or leave, when it is lacking)  an organization. This has direct effects on organizations' capacity to deliver services and software products. 

The past several years have seen dramatic changes in the way people work, driven in large part by the Covid-19 pandemic and the resulting lockdowns, forcing people to work from home. There are numerous studies on how this has affected people in negative ways (e.g. \cite{ralph2020pandemic}). One important theme in this context is Burnout, which is the result of continuous exposure to unhealthy levels of stress. However, there is a paucity of research in software engineering on this topic. Globant, a major provider of software services that has operations across five continents, is reliant on a healthy workforce to conduct its business. Globant is interested in developing better insights into the various factors that might play a role in employee burnout. Thus, in this paper, we report on a large-scale survey at Globant with a primary focus has been on Burnout and its antecedents. 

In particular, we sought to understand the role of organizational culture in relation to Burnout. Organizational culture has been shown to be an important factor in the performance of employees and teams, including in software delivery teams \cite{forsgren2018accelerate}. 
Our theoretical argument in this paper is that all employees within an organization are exposed to the same organizational culture; while they will experience this differently, we believe other factors play a role in why people might experience burnout. In particular, we looked at three factors: people's Sense of Belonging, whether or not an organization advocates a Climate for Learning, and people experiencing Inclusiveness. Further, rather than having a direct link to Burnout, we believe that these factors affect employees' Work Satisfaction, and this is a major predictor for people's experienced Burnout.

\review{We used questions based on an Organizational Culture typology that is focused on how organizations process information and behave when things are not going well, bringing together not only culture, but also management style. We analyzed Organizational Culture as a latent variable that included attributes about sharing bad news with no fear, considering failures as learning opportunities, encouraging cross-functional collaboration, welcoming new ideas, sharing responsibilities, and actively seeking information when needing it.}

We also considered the moderating role of national culture, considering the six dimensions identified by Hofstede. Finally, we conducted our analysis for the whole sample and conducted two different multi-group analyses, distinguishing respondents by gender, and whether or not they are in a leadership role. 

Before we discuss the implications of our results, we discuss a number of limitations of this study that should be kept in mind while interpreting the findings.

\subsection{Threats to Validity}


\subsubsection{Construct Validity}
We adopted and tailored existing measurement instruments when possible, and developed measurement instruments for some constructs based on prior literature. Our analysis of the measurement model confirmed that our constructs were internally consistent, and scored well on convergent and discriminant validity tests. We defined new a construct called Work Satisfaction that included hedonism and satisfaction towards the team and the company. We acknowledge the fact that Burnout is a construct that can be measured by more complex instruments \cite{almen2021reliability}. However, many existing instruments contain a large number of items (questions), which would be impractical in organizational settings because this would negatively affect the response rate.  
\review{In this study we have used respondents' country of residence as a proxy for Power Distance as a dimension of culture as defined by Hofstede \cite{hofstede2011dimensionalizing}. While also used in other studies \cite{cortina2017school}, we acknowledge it is an approximation and not a perfect measure. One potential issue is that we do not know how \textit{long} respondents have lived in their current country of residence. Another potential issue is that contributors' original culture that they grew up with may differ from the culture they now live in. This is why we report the metric as being surrounded by a specific culture instead of having a specific culture. Measuring culture in a more precise way is an important avenue for future work in general.}

\subsubsection{Internal Validity}
We propose a series of hypotheses as associations between different constructs rather than causal relationships, as the present study is a sample study, rather than an experimental study \cite{stol2018abc}. 
Our overall argument is that employees who perceive their organization to have what Westrum \cite{westrum2004typology} labeled a Generative Organizational Culture are satisfied and tend to experience levels of burnout; this line of reasoning is easier to theoretically justify than the suggestion that Burnout leads to a negative organizational culture (what Westrum referred to as a Pathological organization) \cite{westrum2004typology}. 
Further, it is likely that other factors are at play. The coefficient of determination (\textit{R}\textsuperscript{2}) of the endogenous variables ranged between .2 and .4 which in the software engineering context can be considered reasonable. Thus, these results represent a useful starting point for future studies.

\review{Respondents are current employees and we did not collect data from past employees. As the company does not offer the same questionnaire to people who are leaving, we would not have the same data to compare the perspectives of current and past employees. The relationship between burnout and intentions to leave, and also the actual act of leaving are of interest for future work.}

\subsubsection{External Validity}
This survey was conducted within Globant. The response rate numbers are aligned with the overall distribution of the company and therefore can be generalized across the company. Globant is a multi-national company with more than 25,000 employees working in different national cultures. The responses were sufficiently consistent to find full or partial empirical support for the hypotheses. Additional studies that replicate our findings in other companies can further bolster our results.

\subsection{Implications of results}

Our analysis highlights several key findings and implications. In the following, we discuss the supported hypotheses.

\textbf{H1. Generative Organizational Culture $\rightarrow$ Belonging:} 
Our results align with previous research that showed a sense of belonging emerges from a people-centered culture \cite{fawcett2008spirituality} and that an openness to innovation and shared responsibility helps to develop organizational belonging \cite{tabatabaee2016study}. When working in a team that welcomes new ideas, fosters collaboration, and shares responsibilities \cite{westrum2004typology}, people understand how their work contributes toward a common goal leading to affective commitment to the team. Our results showed there was no significant difference between the groups for H1, which proposed a positive association between a Generative Organizational Culture and a Sense of Belonging, as shown in Table \ref{tab:mga_gender}.


\textbf{H2. Generative Organizational Culture $\rightarrow$ Climate for Learning:}
In a generative organizational culture, people do not fear failure because they are trained to learn from mistakes \cite{westrum2004typology}; learning is a key point, considered essential and an investment. An inspiring culture that encourages and enables employees to bring their best efforts and ideas to the team promotes a Climate for Learning \cite{fawcett2008spirituality}. We found that a generative organizational culture is positively associated with Climate for Learning and that there is no significant difference between different groups (either in terms of gender or whether or not respondents fulfilled a leadership role) for this association. 

\textbf{H3. Generative Organizational Culture $\rightarrow$ Inclusiveness:} 
Although the association between Organizational Culture and Inclusiveness is not significantly different according to the leadership position, it is stronger for women ({\textit{B}}=.53) than for men ({\textit{B}}=.41). This difference opens a path to discuss what brings Inclusiveness for minority and majority groups. A lot has been researched about providing safe place for diversity so that everyone feels equally welcomed. However, results shed light that different factors bring the feeling of being included for minority and majority groups.

Hypotheses 1-3 show the benefits of a generative organizational culture where employees have the psychological safety to talk about failures and present new ideas. Therefore, companies should reflect on their team and leadership culture to promote the ideas of generative organizational culture.

\review{Changing the way people behave and work changes culture. Teams can identify helpful practices to create a generative culture that fosters information flow and trust by examining the aspects of Westrum's model of organizational culture \cite{westrum2004typology}, focusing on those behaviors seen in the generative culture:
\begin{itemize}
    \item Cooperation and bridging. Break down silos and create cross-functional teams that include representatives from each functional area of the software delivery process, so everyone shares the responsibility for the software delivery life-cycle. Encourage informal meetings between people who do not understand (or are frustrated by) each other's work. Ask them to understand each other why they do what they do--and invite people to come up with new ideas together.
    \item Train the messengers and let failure lead to inquiry. People must be able to take risks and feel safe to fail, and also to bring bad news without fear in order to make improvements. Hold blameless postmortems, so teams surface problems as early as possible, and solve them more effectively. Instead of blaming, ask questions about the root-cause of failures, in order to improve technical systems, processes, and the organizational culture.
    \item Share risks and responsibilities. Quality, availability, reliability, and security should be everyone's job. One practical example can be ensuring that developers share responsibility for maintaining their code in production.
    \item Encourage novelty. Encouraging employees to explore new ideas can lead to great outcomes. One example of this practice can be giving people time each week for experimentation, hosting internal hackathons and conferences to share ideas and collaborate. When releasing employees from habitual pathways and repetitive tasks, they can be creative, bringing new ideas for processes and products.
\end{itemize}
}
\textbf{H4. Sense of Belonging $\rightarrow$ Work Satisfaction:}
A Sense of Belonging is a human need \cite{baumeister2017need,allen2020psychology,allen2021belonging}. Although it has several different antecedents for people, employees from different groups showed higher levels of Work Satisfaction when they feel they are part of a team. Our analysis based on groups showed similar path coefficients for the association between Sense of Belonging and Work Satisfaction, as we present in Table~\ref{tab:mga_gender}. This indicates that strategies that focus on making employees feel part of the team or the company, pay off, because this positively influences satisfaction, regardless of the group to which the developers belong. Therefore, companies could invest in cohort building, creating opportunities where developers can socialize and develop an emotional connection. Additionally, belonging can be fostered via a team culture where individuals' contributions are appreciated and they can see how their work fits the team's overall goals.

\textbf{H5. Climate for Learning $\rightarrow$ Work Satisfaction:} 
The association between Climate for Learning and Work Satisfaction is significantly different between leaders and non-leaders. While those not in leadership positions are satisfied by having the Climate for Learning, the association was not significant for leaders. There is also no difference when we group developers by their gender. 
Based on these findings, in order to keep employees satisfied in their work, we recommend that companies offer professional development opportunities, where employees can learn new technology and management skills needed to advance their careers.  

\textbf{H6. Inclusiveness $\rightarrow$ Work Satisfaction:}
The association between Inclusiveness and Work Satisfaction holds positively across the whole sample. Additionally, the association showed different strengths when we compared genders. Women are twice more satisfied ({\textit{B}}=.20) when having a welcoming and safe space for diversity than men ({\textit{B}}=.10). Women represent one of the gender minorities and face prejudice and challenges in the tech workplace. A welcoming and safe space allows them to thrive. Companies should evaluate the gender diversity of their tech workforce, expending effort in diversity recruitment and hiring, as well as in training programs that instill diversity and inclusion principles in their teams.

\textbf{H7. Work Satisfaction$\rightarrow$Burnout:}
Work Satisfaction has a negative association with Burnout, with no significant difference across groups.
Burnout is the exhaustion caused by excessive and prolonged workplace stress, which can happen in software delivery teams due to the pressure of deadlines and high performance. However, we showed that Work Satisfaction represents an alleviating factor in reducing Burnout, which is aligned with previous research in software delivery teams \cite{forsgren2021space}. So, although there were different antecedents to Work Satisfaction when achieved, satisfaction reduces Burnout across the groups of gender and leadership positions. 

\textbf{H8.c. Masculinity \texttimes{} Work Satisfaction $\rightarrow$ Burnout}
In a competitive and materialist culture (high levels of Masculinity), there was a great impact of work satisfaction in reducing burnout (This moderation had no effect for women.) \review{According to Hofstede's framework of national cultures \cite{hofstede2011dimensionalizing}}, in Masculine cultures, men ``should be'' and women ``may be'' ambitious, work prevails over family, there is an admiration for the strong, fathers deal with facts and mothers with feelings, and girls cry and boys do not. In feminine cultures, fulfilling multiple social roles without social judgment is encouraged, so both men and women receive cultural support for prioritizing family time over time spent on the job \cite{hofstede2011dimensionalizing}. Finding differences between groups is aligned with previous research that showed that people's well-being is achieved according to their current specific needs \cite{janicijevic2018influence}. The result can be interpreted as masculine cultures drive men to strive for achievement and success, and when they perceive that they are successful (satisfied with their work) it reduces the perception of burnout. In feminine societies, men might feel uncomfortable and burn out more. Another possible interpretation is that men can take other measures to avoid burnout in masculine cultures, because the visible expressions of stress behavior may threaten the masculine value of heroism \cite{tavel2022dispositional}.
Therefore, organizations should consider the national culture where they operate and structure their incentives and career advancement opportunities accordingly. 

\section{Conclusion}
Attention to human factors is critical to software development employees' ability to perform. Globant is a large software services organization whose management sought to understand the concept of Burnout among their workforce. In this paper, we report on a theoretical model that seeks to explain how organizational culture and burnout in software delivery teams are associated. A large-scale survey with over 3,000 respondents provided sufficient data to test this model, and to distinguish between different subgroups (i.e., men/women and people on leadership/non-leadership roles). 
We argue that, given the international nature of this study that also considers the role of national culture (according to the Hofstede 6-D framework), albeit at one company, these findings are of interest to other large multinational organizations. Additionally, there are clear extension points of our study, as well as opportunities to replicate this study, which we think can contribute to a body of knowledge that considers critical human factors such as Burnout.

\section*{Acknowledgments}
We thank all the survey participants for their time and insights. This work is partially supported by the National Science Foundation under Grant Numbers 1815486, 1815503, 1900903, and 1901031, and Science Foundation Ireland grants no. 13/RC/2094-P2 to Lero, the SFI Research Centre for Software, and no. 15/SIRG/3293.

\bibliographystyle{IEEEtran}
\bibliography{reference}

\end{document}